\documentclass[11pt,letterpaper,notoc]{JHEP}
\input epsf.tex
\title{Lower Bound on the Propagation Speed of Gravity 
	from Gravitational Cherenkov Radiation}
\author{Guy D.~Moore and Ann E.~Nelson\\
	Department of Physics, University of Washington \\ Seattle,
	Washington, WA 98195, USA.}
\preprint{UW/PT 01-18}
\abstract{
Recently, interesting 4-D Lorentz violating models have been
proposed, in which all particles have a common maximum velocity $c$, but
gravity propagates (in the preferred frame) with a different maximum
velocity $c_g \neq c$.  We show that the case $c_g < c$ is very tightly
constrained by the observation of the highest energy cosmic rays.
Assuming a galactic origin for the cosmic rays gives a conservative
bound of $c-c_g < 2 \times 10^{-15} c$; if the cosmic rays have an
extragalactic origin the bound is orders of magnitude tighter, of order
$c-c_g < 2 \times 10^{-19} c$.}

\def\lsim{\mbox{~{\raisebox{0.4ex}{$<$}}\hspace{-1.1em}
	{\raisebox{-0.6ex}{$\sim$}}~}}
\def\k{{\bf k}}
\def\p{{\bf p}}

\def\be{\begin{equation}}
\def\ee{\end{equation}}

\newcommand{\rem}[1]{}

\newcommand\pcite[1]{\protect{\cite{#1}}}

\begin{document}

\section{Introduction and Motivation}

A new generation of  experiments  utilizing long baseline
interferometers is commencing  to search for 
gravitational waves 
\cite{Abramovici:1992ah,Giazotto:1990gw,Schutz:1999wt,Tagoshi:2000bz}.%
\footnote{for
         reviews, see ref. \pcite{Thorne:1995xs,Thorne:1997ut,Will:2001mx}.}
Within ten years or so it is quite possible that gravitational
waves from astrophysical sources  will have been observed. As well as
opening a new astronomical and cosmological window, this may
allow for  tests   of unprecedented accuracy of the general 
relativistic prediction for the
speed and polarization of gravitational waves. Accurate measurement of
the speed of propagation of gravitational waves 
can constrain  extra-dimensional
``brane-world'' theories in which gravity
propagates in  the bulk of  extra dimensions
\cite{Akama:1982jy,Holdom:1983jh,Rubakov:1983bb,Arkani-Hamed:1998rs,%
Antoniadis:1998ig,Randall:1999ee,Randall:1999vf},
while the  particles of the Standard Model are confined to a 3+1 dimensional
subspace known as a brane. It has been
argued that  in many cases Poincar\'e invariance should be 
violated in the bulk
\cite{Chung:1999zs,Ishihara:2000nf,Csaki:2000dm,%
Hebecker:2001nv,Moffat:2001hz,Youm:2001zk,Caldwell:2001ja}
giving rise to an anomalous dispersion relation for
gravitational waves.%
\footnote{
	In ref. \pcite{Csaki:2000dm} it was argued on the basis of
	examining geodesics that in such Poincar\'e non-invariant warped
	geometries gravity would always travel at least as fast as light. The
	argument is that  if the bulk velocity were slower than light,
	gravity could just stay on the brane.  Such a geodesic analysis
	actually establishes the maximum propagation speed over all
	Kaluza-Klein states of the graviton with any wave number, rather
	than the propagation speed of the massless graviton.  The constraint
	we find in this paper applies if the graviton or {\em any} KK
	state propagates with subluminal phase velocity when its
	energy is less than that of the most energetic cosmic rays.
	}
No Lorentz violation would  show up
in the standard model, provided our brane is Poincar\'e invariant.

 The ADS/CFT correspondence \cite{Maldacena:1998re} has given rise to
a purely 4 dimensional interpretation, in the infrared, 
of warped  higher dimensional
geometry
\cite{Verlinde:1999fy,Witten99,Gubser:1999vj,%
Arkani-Hamed:2000ds,Rattazzi:2000hs,Hebecker:2001nv}.
It is conjectured that a  bulk with 4-dimensional
Poincar\'e invariance and 5-dimensional anti-deSitter (AdS) invariance
may be given a dual 4-dimensional interpretation as a
dark `conformal sector' which interacts
gravitationally. A less symmetric bulk may be regarded simply as
deviations of the dark sector from conformality and/or Poincar\'e
invariance. Deviations from Poincar\'e invariance would be expected if
the conformal sector
 were at finite
temperature. If gravitational deviations from Poincar\'e invariance
can be interpreted simply as a modified dispersion relation for
gravitational waves due to interaction with a hot dark sector, then 
gravitational waves always travel {\it slower} than
light \cite{Peters:1974gj}.%
\footnote{In this case, 
if we enforce the additional constraint that the energy density off the
brane is less than the critical density 
$\rho_{\rm crit} = 8\pi G_{\rm N} H_0^2 /3$, as seems necessary on
cosmological grounds, then the propagation velocity is strongly
constrained.  On wavelengths longer than the compactification radius,
the gravitational waves will have the dispersion relations of 4-D
gravitational waves interacting with (off-brane) dark matter.  If the
dark matter behaves as dust, the frequency $\omega$ dependent
propagation speed is 
$(c-c_g)/c \simeq 2\pi G \rho / \omega^2$ \protect{\cite{Peters:1974gj}}; for
other equations of state the coefficient will differ but the form should
be the same.  If the compactification radius is, say, 1mm, the velocity
difference for wavelength of order the compactification radius
is $(c-c_g)/c < 10^{-58}$.
}

In this note we remark that if gravity is slower than light, one
expects particles moving faster than the speed of gravity to emit
``gravi-Cherenkov radiation,'' in analogy with the  Cherenkov radiation
emitted by particles moving faster than  light in a medium. The
existence of high energy cosmic rays which have travelled from
astronomical distances without losing all their energy to gravi-Cherenkov
radiation places a
strong lower bound on the speed of gravitational waves 
with very short wave lengths (energies of order $10^{10}$GeV).

We begin by reviewing Cherenkov radiation from a particle physics point
of view.  Then we generalize to gravitational Cherenkov radiation, and 
apply it to bound the speed of propagation of gravitational waves.
Similar limits have been placed on Lorentz violating effects in which
different standard model particles have different maximum propagation
velocities, by considering ordinary Cherenkov radiation 
\cite{Coleman:1998ti,Stecker:2001vb}.  Our constraints will turn out to
be weaker because the efficiency of gravitational Cherenkov radiation is
orders of magnitude less than the efficiency of ordinary Cherenkov
radiation.  

\section{Cherenkov radiation}

Cherenkov radiation occurs when a charged particle moves faster than the
local (in-medium) speed of light.  Intuitively, Cherenkov radiation
happens because 
the charged particle ``outruns'' its own electromagnetic field.  It is 
easiest to see that this leads to energy loss by considering the case
where the speed of electromagnetic propagation is zero; then the
electric flux of a charge will trail out along its
past trajectory.  As the charge moves, it lengthens the trajectory and
pays the energy cost of the added electric field energy.

In normal applications Cherenkov radiation occurs when some medium
effect causes the propagation speed of electromagnetic radiation to be
less than light speed, for light with frequencies below some cutoff
frequency which is typically set by an atomic 
excitation frequency.  In this case the index of refraction varies with
frequency in an important way, and Cherenkov radiation only occurs for
frequencies much less than the energy of the relativistic charged
particle.  Therefore a classical treatment is applicable.
For simplicity, however, we will consider cases where the
index of refraction $n$ is frequency independent in the deep
ultraviolet.  We also only need
treat cases where $c_g$ is
very close to the maximum propagation speed of the
particle, that is, the index of refraction $n = c_{\rm particles}/c_{\rm
light} = 1 + \epsilon$ with $\epsilon \ll 1$ a constant.  Cherenkov radiation
corresponds to the process shown in Fig.~\ref{fig1}.  
\FIGURE{%\begin{figure}[t]
\vspace{0.1in}
\centerline{\epsfxsize=2in\epsfbox{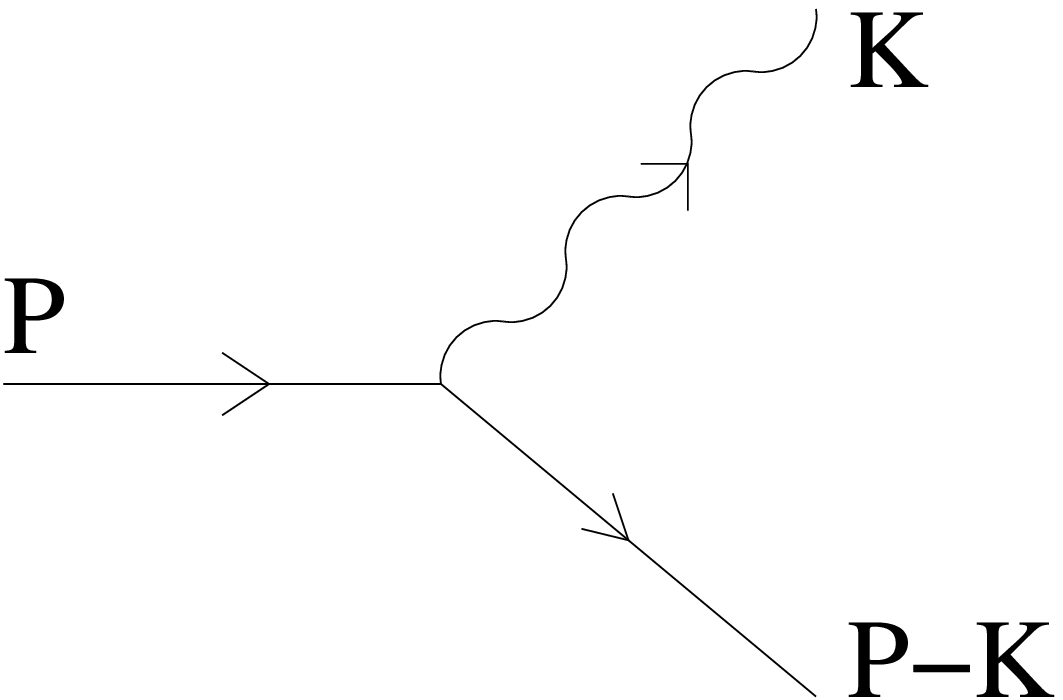}}
\vspace{-0.2in}
\caption{\label{fig1} Diagram responsible for Cherenkov radiation.}}
%\end{figure}
Normally this
process is kinematically forbidden and we would never consider it.
However, the dispersion relation for a photon with a constant index of
refraction $n>1$ is $|\k| = n k_0$.  Therefore, for $n>1$ a photon
carries more momentum than energy,
which makes the process kinematically allowed for sufficiently
relativistic charge carriers.  Note that it is the phase velocity of the
light which is important here, not the group velocity.

Evaluating the two-body
decay rate by standard techniques, and approximating $n\simeq 1$
whenever it is safe to do so, gives a photon emission rate
from a scalar of charge $e$, mass $m$ and momentum $\p$ of
\begin{equation}
\frac{dn_\gamma}{dt} = \frac{1}{2p_0} \int \frac{d^3 \k \, dk_0}{(2\pi)^4}
	\: 2\pi \delta \! \left[n^2 k_0^2-\k^2\right] \: 
	2\pi \delta \! \left[(p_0{-}k_0)^2 - (\p{-}\k)^2-m^2 \right] 
	\; e^2 \sum_\epsilon \left( (2P-K)^\mu 
	\epsilon_\mu \right)^2 \, ,
\end{equation}
where the sum is over the two transverse photon polarization states
$\epsilon$.  The first delta function forces the photon to be on the
(modified) mass shell $|\k| = n k_0$ and the second is the mass shell
condition for the outgoing scalar, while the final factor is the square
of the matrix element.  Writing $|\p| \equiv p$ and $|\k|
\equiv k$, and defining $\theta$ to be the angle between $\p$ and $\k$,
$\p \cdot \k = pk \cos \theta$, the second delta function can be
rewritten as
\begin{eqnarray}
\delta \! \left[(p_0{-}k_0)^2 - (\p{-}\k)^2-m^2 \right]
	& = & \delta \left[ 2pk \cos \theta - 
	2 p_0 k_0 - k^2+k_0^2 \right]
	\nonumber \\
	& = & \frac{1}{2pk} \delta \left[ \cos \theta - 
	\frac{1}{\beta n} - \frac{(n^2-1)k}{2n^2 p} \right] \, ,
\end{eqnarray}
where we have written the propagation velocity of the incoming charged
particle as $p/p_0 \equiv \beta \simeq 1$.  The photon emission rate
becomes 
\begin{equation}
\label{eq:photo_result}
\frac{dn_\gamma}{dt} = \frac{e^2}{4\pi} \int_0^{k_{\rm max}} dk 
	\sin^2 \theta \, , 
	\qquad
	\cos \theta = \frac{1}{\beta n} + \frac{(n^2-1)k}{2n^2 p}
	\, , \qquad
	k_{\rm max} = p \: \frac{(\beta n)-1}{\beta(n-1)} \, ,
\end{equation}
where, note, we have already approximated $(n{-}1) \ll 1$ and
$(1{-}\beta)\ll 1$.
The small $k$ limit is the same as
the classical electromagnetic result.  The power emitted is obtained by
inserting a factor of $k$ in the integrand.  Note that, for $k \sim p$,
spin dependent corrections become $O(1)$.

\section{Gravitational Cherenkov radiation}

The rate of energy loss by gravitational Cherenkov radiation is very
similar.  There are a few key physical differences, however.
Rather than the electric charge $e^2$, it is the gravitational constant
times the energy squared, $G_{\rm N}\, p^2$, which will appear as the
prefactor.  Also, the tensor nature of the gravitational interaction
reduces emission at small opening angle, so it is $\sin^4 \theta$,
rather than $\sin^2 \theta$, which appears.  The same diagram,
Fig.~\ref{fig1}, is responsible for gravitational Cherenkov radiation,
with the wavy line now representing a graviton rather than a photon.
The relevant Feynman rules can be obtained from Appendix A of
\cite{Han:1998sg}.\footnote
	{%
	The Feynman rules in \pcite{Han:1998sg} are for arbitrary KK states 
	of the graviton and must be truncated for our purposes; only
	the spin $\pm 2$ polarization states in Eq.~(A.3) of that 
	paper are to be summed over, and indices referring to KK state
	number are to be ignored.
	}
The rate of
graviton emission by gravitational Cherenkov radiation, again taking
$(n{-}1) \ll 1$ and $(1{-}\beta) \ll 1$, is
\begin{eqnarray}
\frac{dn_{\rm grav}}{dt} 
	& = & \frac{1}{2p} \int \frac{d^3 \k d k_0}{(2\pi)^4} \:
	\: 2\pi \delta \! \left[n^2 k_0^2-\k^2\right] \:
	2\pi \delta \! \left[(p_0{-}k_0)^2 - (\p{-}\k)^2-m^2 \right] 
	\times \nonumber \\ && \qquad \qquad \times 
	\, 16 \pi G_{\rm N}\, \sum_\epsilon \left( \epsilon^{\mu \nu}
	P_\mu (P{-}K)_\nu \right)^2 \, ,
\end{eqnarray}
where the sum is over the two transverse traceless graviton
polarizations, which project out the components of $P_\mu$,
$(P{-}K)_\mu$ orthogonal to
$K_\mu$.  The result for energy loss per unit time is obtained by
multiplying the integrand by $k$, and is
\begin{equation}
\frac{dE}{dt} \simeq 
G_{\rm N}\, p^2 \int_0^{k_{\rm max}} k dk \sin^4 \theta \, ,
\end{equation}
where $k_{\rm max}$ and $\theta$ are the same as in
Eq.~(\ref{eq:photo_result}).  If we take the limit $(1{-}\beta) \ll
(n{-}1)$, which will be appropriate to our application, then the
simpler expressions $k_{\rm max} \simeq p$ and 
$\sin^2 \theta \simeq \theta^2 \simeq 2(n{-}1)(1{-}k/p)$ hold, and the
integral is easy;%
\footnote{
    A result similar to Eq.~\ref{eq:result} has been derived by Pardy
    \cite{Pardy:1994dv}, for general $\beta<1$ and $n>1$ but using a 
    classical technique only applicable for $k \ll p$.  After correcting
    an error in Eq.~(19) there [his $(n^2/\beta^2 + 1)^2$ should read
    $(n^2 / \beta^2 - 1)^2$] his answer agrees with ours in the overlapping
    domain of validity, $k\ll p$ and $(1{-}\beta) \ll (n{-}1) \ll 1$.
    }
\begin{equation}
\frac{dE}{dt} \simeq 4 G_{\rm N}\, (n{-}1)^2
\int_0^p (p-k)^2 k dk = 
\frac{G_{\rm N}\, p^4 (n{-}1)^2}{3} \, .
\label{eq:result}
\end{equation}
The energy loss arises predominantly by shedding gravitons with energy
$k\lsim p/2$.  We have presented this calculation
for scalars, but we have checked that the result is the same for fermions.  

Integrating this equation, the relation between the travel time, the
initial momentum $p_{\rm init}$, and the final momentum $p$ is
\begin{equation}
t_{\rm travel} = \frac{1}{G_{\rm N}(n{-}1)^2} \left( \frac{1}{p^3} 
	- \frac{1}{p_{\rm init}^3} \right) \, .
\label{eq:tmax}
\end{equation}
Therefore, a particle of momentum $p$ cannot possibly have been
traveling for longer than $t_{\rm max} = m_{\rm pl}^2/(n-1)^2 p^3$.

\section{Application to cosmic rays}

Several cosmic rays have been observed with energies in excess of
$10^{11}$GeV \cite{Nagano:2000ve}.
The highest energy cosmic ray which has been observed was probably a
proton, of energy $\sim 3 \times 10^{11}$GeV \cite{Bird:1995uy}.  A proton
is a composite object.  Viewed at the energy scale $k \sin \theta \sim
10^3$GeV, it is made up of pointlike partons with typical momentum
fraction $x \sim 1/10$.  Let us assume conservatively that the proton
arrived at Earth from a distance of order the distance to the galactic
center, say 
$\sim 10$kpc (Kiloparsec).  In particle physics units,
$10 {\rm kpc} \simeq 1.57\times 10^{36} {\rm GeV}^{-1}$.  Using the
bound on the time of flight, Eq.~(\ref{eq:tmax}), we find
\begin{equation}
n{-}1 \leq \sqrt{ \frac{m_{\rm pl}^2}{(0.1 E)^3 t}}
	\simeq 2 \times 10^{-15} \, .
\end{equation}
Such an $n$ indeed satisfies $(1{-}\beta) \simeq m^2/2 E^2 
\sim 10^{-23} \ll (n{-}1)
\ll 1$, so our analysis is self-consistent.  This bound is strong enough
that coincidence tests between gravitational wave detectors and other
observations (gamma rays, etc.) of distant gravity wave sources are
unlikely to do better.

Our result is also very conservative.  For instance, we neglect that
the emission of a
graviton with transverse momentum $k \sin \theta \sim 100$GeV would
break up the proton, so the actual energy loss to the proton would be
substantially more than the energy of the graviton.  Also, in many
higher dimensional theories there will be KK states of the graviton
with mass $\lsim 100$GeV.  These 
will also satisfy $|\k|>k_0$ at energies $k_0 \sim 10^{10}$GeV, so the
proton will radiate them as well, increasing the efficiency of the
energy loss mechanism.  For instance, if there is one extra millimeter
scale dimension, our bound strengthens by $\sim 7$ orders of magnitude.

Our bound also assumes that there be some exotic physics which can
generate the highest energy cosmic rays within the galactic halo.
The result tightens if we do not
assume such exotic physics.  This is an experimental question which will
be resolved when the angular
distribution on the sky of the highest energy cosmic rays is determined
\cite{Dubovsky:1998ru}.  Consider instead the possibility
that the highest energy
cosmic rays are produced by the ``Z-burst'' mechanism, in which very
high energy neutrinos produced at cosmological distances annihilate with
relic neutrinos via the $Z$ boson resonance
\cite{Weiler:1982qy,Weiler:1997sh}.  In this
case, the energy loss of the high energy
neutrino sets a limit on $n{-}1$ which is 
much tighter.  The energy of the neutrino must exceed the energy of the
cosmic ray primary, $\sim 3\times
10^{11}$GeV, after traveling a cosmological distance, say $\sim 2 {\rm
Gpc} \sim 3 \times 10^{41}{\rm GeV}^{-1}$.  In this case the bound
becomes 
\begin{equation}
n{-}1 \leq 1.3 \times 10^{-19} \, .
\end{equation}
This bound is much tighter but is model dependent.

We can say less about the case $n<1$, where gravity propagates
faster than the speed of light.  In this case gravitational Cherenkov
radiation is impossible.  On the other hand, 
in terms of the particle physics Lorentz
metric, a graviton has a timelike energy 4-vector, so it is
kinematically allowed for a graviton to convert into, say, two photons.
However, as we have seen, the reaction rate is very highly energy
dependent, so the relatively low frequency gravity waves we expect from
most gravity wave producing phenomena are completely unaffected.
Therefore the direct observational bounds on $n<1$ are set by precision
tests of gravity, and are very much weaker than for $n>1$.

\bibliography{cherenkov} 
\bibliographystyle{JHEP} 

\end{document}